\documentclass[reprint, amsmath, amssymb, aps, showkeys]{revtex4-2}

\usepackage{gensymb}
\usepackage{booktabs}
\usepackage{graphicx}
\graphicspath{{./Figures/}}
\usepackage{multirow}
\usepackage{dcolumn}
\usepackage{bm}
\usepackage{times}
\usepackage{algorithm}
\usepackage{algorithmicx}
\usepackage{algpseudocode}
\usepackage{listings}
\usepackage{natbib}
\usepackage{placeins}
\usepackage{array}
\usepackage[explicit]{titlesec}
\usepackage{color}
\usepackage{mdframed}
\usepackage{multirow}
\usepackage{booktabs} 

\usepackage[colorlinks = true,
linkcolor = blue,
urlcolor  = blue,
citecolor = blue,
anchorcolor = blue]{hyperref}

\newcommand{\Tx}{T_\text{x}}
\newcommand{\Tg}{T_\text{g}}
\newcommand{\Tl}{T_\text{l}}

\begin{document}

\preprint{APS/123-QED}

\title{A Multi-agent Framework for Physical Laws Discovery}

\author{Bo Hu$^1$}
\author{Siyu Liu$^{1,2}$}
\author{Beilin Ye$^1$}
\author{Yun Hao$^3$}
\author{Yanhui Liu$^4$}
\author{Yang Lu$^{1,2}$}
\author{Ju Li$^{5,6}$}
\author{David J. Srolovitz$^{1,2}$}
\author{Tongqi Wen$^{1,2}$}

\affiliation{$^1$Center for Structural Materials, Department of Mechanical Engineering, The University of Hong Kong, Hong Kong, China}
\affiliation{$^2$Materials Innovation Institute for Life Sciences and Energy (MILES), HKU-SIRI, Shenzhen, China}
\affiliation{$^3$School of Software Engineering, South China University of Technology, Guangzhou, China}
\affiliation{$^4$Institute of Physics, Chinese Academy of Sciences, Beijing, China}
\affiliation{$^5$Department of Materials Science and Engineering, Massachusetts Institute of Technology, Cambridge, MA, USA}
\affiliation{$^6$Department of Nuclear Science and Engineering, Massachusetts Institute of Technology, Cambridge, MA, USA}

\date{\today}

\begin{abstract}
Discovering explicit physical laws has traditionally depended on human intuition and domain expertise. Recent advances in artificial intelligence, particularly large language models (LLMs), offer a new route to accelerate this process by automating key steps from hypothesis generation to interpretable model construction. Here we develop an LLM-based multi-agent framework for physical-law discovery that integrates literature-guided variable selection, hypothesis formulation, symbolic regression, formula derivation, and mechanistic explanation. We validate the framework on three representative materials problems: the glass-forming ability (GFA) of metallic glasses, the Vickers hardness of compounds, and the Young's modulus of multi-component alloys. Using physically and chemically meaningful descriptors as inputs, the discovered formulas achieve strong agreement with reference data, with correlation coefficients up to 0.94 (GFA), 0.86 (hardness), and 0.94 (Young's modulus), while remaining compact and interpretable. Beyond fitting, the Young's modulus formula generalizes to quaternary and quinary alloys, improving prediction accuracy by up to 78.8\% relative to the classical rule of mixtures. By integrating cross-disciplinary knowledge, reflection mechanisms, and expert-like reasoning ability into symbolic regression, our AI-centric framework offers a robust and extensible platform for automated physical laws discovery, demonstrating that AI can increasingly serve as an essential role in modern scientific research by thinking and acting like field experts.

\end{abstract}

\maketitle

Artificial intelligence (AI), especially machine learning and deep learning, is reshaping scientific research by accelerating the extraction of patterns and predictive relations from large-scale data. Data-driven models have achieved notable advances in diverse domains, including the simulation of quantum systems~\cite{liDeeplearningDensityFunctional2022}, the prediction of materials properties~\cite{taoMachineLearningPerovskite2021,jhaElemNetDeepLearning2018a}, and the generative design of new crystal structures~\cite{zeniGenerativeModelInorganic2025}. In materials science, these developments are further enabled by the rapid growth of open databases, such as the Materials Project~\cite{jainCommentaryMaterialsProject2013}, AFLOW~\cite{curtaroloAFLOWAutomaticFramework2012,gossettAFLOWMLRESTfulAPI2018}, ICSD~\cite{zagoracRecentDevelopmentsInorganic2019}, and Materials Cloud~\cite{yangMatCloudHighthroughputComputational2018}, which support systematic exploration of structure-property relationships. Nevertheless, two fundamental limitations constrain purely data-driven deep learning for scientific discovery. (i) Model performance depends critically on the availability of high-quality labelled data, yet reliable experimental measurements remain scarce because they are expensive and time-consuming to obtain. (ii) Many deep networks provide accurate predictions without interpretable mechanisms, functioning as ``black boxes'' that hinder physical understanding and the derivation of generalizable principles. Overcoming these barriers calls for explainable AI approaches that retain predictive power while yielding transparent, physically meaningful relationships.

Beyond predictive modeling, AI is increasingly being explored for scientific reasoning and discovery. Recent advances in large language models (LLMs) have produced systems that can solve complex tasks spanning programming, mathematical reasoning, and technical writing, motivating a central question: can LLMs move from being assistive tools to functioning as ``virtual scientists''? Owing to their broad domain knowledge and emerging reasoning capabilities, LLMs are being explored for key components of the scientific workflow, including hypothesis generation, literature synthesis, experimental planning, and interpretation and communication of results. Early demonstrations suggest that such models can be coupled with automated platforms to support materials and chemistry experimentation and accelerate the discovery of new materials~\cite{zhang_2025_nature,niMatPilotLLMenabledAI2024}. However, current LLMs remain prone to hallucinations and unverified claims, which poses a serious barrier to reliable scientific use. Inspired by the collaborative nature of human research, multi-agent frameworks offer a promising route to improve robustness by enabling iterative debate, cross-checking, and verification among specialized agents, thereby strengthening the credibility of AI-assisted scientific discovery.

Scientific progress is often marked by the discovery of compact physical laws--from Newton's laws of motion to Kepler's laws of planetary dynamics~\cite{udrescuAIFeynmanPhysicsinspired2020}--typically expressed as explicit mathematical equations that expose underlying mechanisms. Such laws not only summarize observations but also enable explanation, extrapolation, and the formation of new conceptual frameworks; indeed, paradigm shifts frequently co-evolve with the articulation of new governing principles~\cite{kuhn1997structure}. Although modern deep learning models can deliver highly accurate predictions, their limited interpretability hampers the extraction of mechanistic insight and generalizable scientific understanding. In contrast, closed-form expressions directly encode relationships among variables and therefore provide a natural basis for interpretation and theory building. Consequently, inferring interpretable equations from experimental and simulation data remains a central challenge in data-driven science. Symbolic regression (SR) has emerged as a leading explainable approach for this purpose, aiming to identify analytical expressions that best describe the dependencies among physical variables~\cite{wangScientificDiscoveryAge2023,angelisArtificialIntelligencePhysical2023}. More recently, LLMs, leveraging extensive pre-training and strong priors over scientific syntax and concepts, have been incorporated to guide and accelerate SR, thereby improving the automated discovery of governing equations~\cite{ICSR,shojaeeLLMSRScientificEquation2024,duLLM4EDLargeLanguage2024}.

In this work, we introduce a general LLM-driven multi-agent framework~\cite{wu2024autogen,NEURIPS2023_a3621ee9} for the automated discovery of physical laws from scientific data. The framework orchestrates multiple agents to emulate core stages of the scientific workflow, including literature review, hypothesis generation, data curation, symbolic regression, and the derivation and interpretation of closed-form equations. Methodologically, our symbolic regression module combines beam search with LLM-based reflection mechanisms~\cite{Reflextion,gouCRITICLargeLanguage2024} to propose, evaluate, and iteratively refine candidate expressions in a structured manner. We evaluate the framework on three representative materials-science problems: predicting the glass-forming ability (GFA) of metallic glasses, the Vickers hardness of compounds~\cite{tantardiniMaterialHardnessDescriptor2024}, and the Young's modulus of multi-principal element alloys (MPEAs)~\cite{vazquezEfficientMachinelearningModel2022a}. Using Gemini-2.5-flash as the base model, the discovered formulas achieve predictive performance of up to $R^2=0.94$, $0.86$, and $0.94$ on these tasks, respectively. Because experimental materials data can be noisy and heterogeneous, predictive metrics alone may not fully reflect the scientific utility of the inferred laws. To further assess practical value and generalization, we apply the discovered Young's modulus equation to the design of previously unseen quaternary and quinary MPEAs. The resulting formula reduces the mean absolute percentage error (MAPE) by up to 78.8\% relative to existing empirical relations, while offering substantially higher computational efficiency than first-principles calculations and foundation atomic models. Together, these results demonstrate that LLM-enabled multi-agent reasoning can bridge data-driven learning and interpretable symbolic modeling, providing a scalable route toward automated, physically meaningful law discovery across scientific domains.

\section*{Framework for physical laws discovery}

\begin{figure*}[ht]
  \centering
 \includegraphics[width=0.95\textwidth]{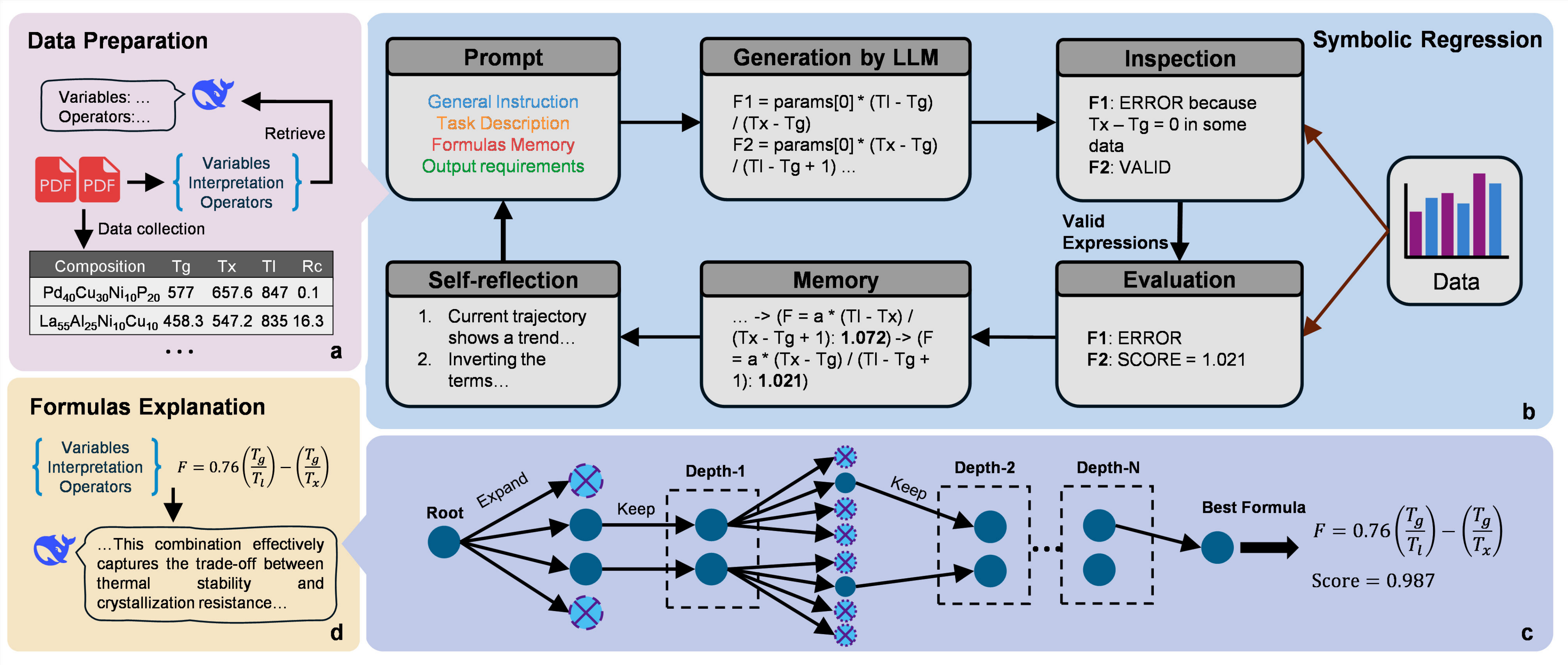}%
 \caption{Schematic of the proposed multi-agent framework for physical-law discovery. (a) A reasoning LLM conducts literature review and supports data preparation. (b) Multi-agent collaboration for proposing, evaluating, and refining candidate formulas with trajectory-based feedback. (c) Beam search balances predictive accuracy and expression complexity by retaining the top-$K$ candidates at each depth. (d) An explanation agent with RAG contextualizes and interprets the discovered formulas.}
 \label{fig.overview}
\end{figure*}

As illustrated in Fig.~\ref{fig.overview}, our framework proceeds through six stages: (i) literature review and data preparation to define the target property, relevant descriptors, and a consistent dataset; (ii) construction of a task-specific prompt that specifies the task objective, input variables, admissible mathematical operators, and evaluation criteria, supplemented with few-shot exemplars; (iii) initialization of the symbolic-regression search by setting beam-search hyperparameters; (iv) generation, retrieval, and scoring of candidate formulas, followed by selection of top-performing expressions; (v) scientific interpretation of the discovered laws to assess plausibility, consistency, and physical meaning; and (vi) deployment of the most accurate and interpretable formulas in downstream scientific tasks, such as prediction and materials design.

During data preparation, we employ an LLM-based agent to recommend candidate variables and admissible mathematical operators for symbolic regression, based on an external scientific knowledge base. Because LLMs are trained on data up to a fixed cutoff and may lack awareness of recent advances, we incorporate retrieval-augmented generation (RAG)~\cite{NEURIPS2020_6b493230} to provide up-to-date, domain-specific context. As shown in Fig.~\ref{fig.overview}(a), once the user specifies a research question, we retrieve relevant publications (or accept user-supplied references) and encode them into a GraphRAG/LightRAG index~\cite{guoLightRAGSimpleFast2025}. This structured repository then functions as an external knowledge base that the agent can query to justify and constrain the SR search space, thereby producing better-informed choices of variables, operators, and plausible equation forms.

In the symbolic regression stage, the search is organized as a tree, where each node corresponds to an iterative multi-agent loop (Fig.~\ref{fig.overview}(b)). At each iteration, a \emph{generation agent} proposes candidate analytical expressions using a customized prompt comprising four components (see Supplementary Information (SI) for details): \textit{General Instruction}, \textit{Task Description}, \textit{Formula Memory}, and \textit{Example Output}. The \textit{General Instruction} specifies syntax constraints and generation rules, the evaluation objective, and safeguards to reduce hallucinated or invalid expressions. The \textit{Task Description} formalizes the user-defined problem, including scientific context, dependent and independent variables, and the allowed operator set. The \textit{Formula Memory} records the current node state and previously explored formulas together with their scores and reflection feedback, which discourages redundant candidates and promotes iterative refinement. Finally, \textit{Example Output} provides few-shot exemplars and enforces a predefined JSON response schema, in which the \texttt{formula} field contains the proposed symbolic expression and the \texttt{theory} field provides a concise theoretical rationale.

All candidate laws are returned as executable Python expressions, which enables automated numerical fitting and evaluation. We assess each formula using a composite score that balances predictive error and symbolic complexity,

\begin{equation}\label{eq.evaluation}
  s(\hat{f}(\boldsymbol{x})|y) = \text{Error}(\hat{f}(\boldsymbol{x}),y) + \lambda\mathcal{C}(\hat{f})
\end{equation}
where $y$ denotes the ground-truth targets, $\hat{f}(\boldsymbol{x})$ is the model prediction, and $\mathcal{C}(\hat{f})$ quantifies the expression complexity. Unless otherwise stated, we use the normalized mean squared error (NMSE) to measure predictive accuracy,
\begin{equation}
  \text{NMSE}(\hat{f}(\boldsymbol{x}),y) = \frac{N-1}{N}\frac{\sum_{i=1}^{N}\left(y_i-\hat{f}(x_i)\right)^2}{\sum_{i=1}^{N}(y_i-\bar{y})^2 + \epsilon}
\end{equation}
where $N$ is the number of samples, $\bar{y}$ is the sample mean, and $\epsilon$ is a small constant (set to $1\times10^{-10}$) to avoid division by zero. In Eq.~\eqref{eq.evaluation}, the first term rewards accuracy, whereas the second penalizes overly complex expressions; $\lambda$ controls the trade-off. We define $\mathcal{C}(\hat{f})$ as a weighted operator count, i.e., the sum over operators of their occurrences multiplied by user-specified weights, allowing different preferences for specific operators. Therefore, a lower score indicates a better-performed formula. To improve search efficiency and generalization, we ask the LLM to avoid explicit numerical constants and to instead introduce free parameters (e.g., \texttt{param[0]}, \texttt{param[1]}). During evaluation, these parameters are optimized to minimize $s(\hat{f}(\boldsymbol{x})|y)$, and the minimized score is used to rank candidates. Expressions that cannot be evaluated or optimized (e.g., due to division by zero or numerical overflow) are assigned a \texttt{None} score and excluded from subsequent analysis.

To efficiently explore the combinatorial space of candidate laws, we employ a beam-search strategy for formula generation and augment it with an explicit memory of past trials. After evaluation, each candidate expression and its score are appended to a trajectory log, e.g., ``(F$_0$=$\cdots$: Score$_0$)$\rightarrow$(F$_1$=$\cdots$: Score$_1$)$\rightarrow\cdots$'' (Fig.~\ref{fig.overview}(b)). This trajectory is fed back to the LLM to maintain context, discourage repeated proposals, and enable systematic refinement. We further incorporate a self-reflection mechanism implemented as an auxiliary \emph{reflection agent}. At each node, the reflection agent inspects the accumulated trajectory, diagnoses common failure modes (e.g., unnecessary complexity or poor fit), and returns actionable guidance for the next generation step. In addition, it functions as a verification layer that flags and filters expressions violating predefined constraints (such as restricted operator sets or invalid syntax), thereby reducing hallucination-induced errors and keeping the search within a physically meaningful hypothesis space.

We use beam search to control the growth of the hypothesis space. Unlike depth-first search (DFS) or breadth-first search (BFS), which may expand exponentially, beam search retains only the top-$K$ partial solutions at each depth, thereby focusing computation on the most promising candidates while preserving diversity. The procedure is parameterized by three hyperparameters: the maximum depth $D$, the number of new formulas proposed per node $N$, and the beam width $K$ (the number of nodes retained at each depth). At depth $d\in\{1,2,\ldots,D\}$, each of the $K$ retained nodes can spawn up to $N$ child candidates, yielding at most $N\times K$ newly evaluated formulas. Each child node (Fig.~\ref{fig.overview}(c)) inherits the search trajectory of its parent and extends it with the newly proposed expression, enabling progressive refinement. After scoring, valid candidates are stored in a formula database, and only the $K$ nodes with the lowest scores are propagated to the next depth. The search terminates at depth $D$, and the best-scoring formula over all explored nodes is returned. To further reduce wall-clock time, candidate generation and evaluation for the $K$ nodes at each depth are executed in parallel.

After selecting the best-performing formulas, providing scientifically grounded interpretations is essential for human understanding and downstream use. We therefore introduce an \emph{Explanation Agent} that leverages the external knowledge base constructed during data preparation to contextualize and rationalize the discovered relations (Fig.~\ref{fig.overview}(d)). Conditioned on the final expression, the relevant variables, and domain references, the agent produces mechanistic and theory-consistent explanations, clarifying plausible physical meanings, assumptions, and limitations. This step complements the numerical evaluation by helping ensure that the resulting expressions are not only computationally accurate, but also scientifically interpretable and actionable for materials research and related disciplines.

\section*{Formula derivation trajectory}
We evaluate the proposed multi-agent framework on three representative materials-science problems to demonstrate its capability and generality: (i) GFA of metallic glasses, (ii) Vickers hardness of inorganic compounds, and (iii) Young's modulus of multi-principal element alloys (MPEAs). The corresponding datasets, input descriptors, and target properties for each task are provided in the Methods section.

 \begin{figure*}[ht]
 	\centering
 	\includegraphics[width=0.95\textwidth]{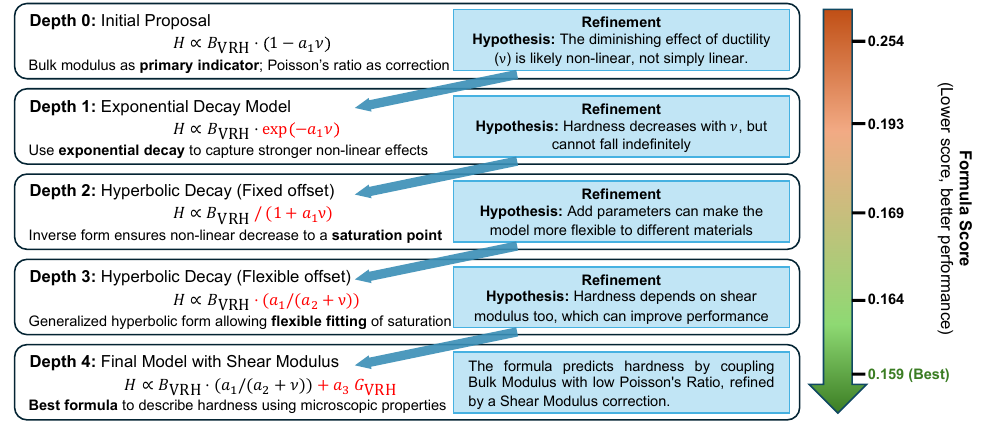}
 	\caption{Evolution trajectory of the best-discovered formula for the hardness prediction task. Using the stored derivation history in the memory module, the agents iteratively propose hypotheses and refine the expression across search depths. The left panel summarizes, at each depth, the structural changes to the formula together with the associated rationale, whereas the arrow on the right highlights the corresponding score improvements (lower is better).}
 	\label{fig.reasoning}
 \end{figure*}
 
 Beam search explores candidate expressions in a tree structure, where each node inherits and extends the derivation trajectory of its parent. Using the hardness-prediction task as an illustrative example, we reconstruct the complete evolution path by tracing the highest-scoring terminal node back to the root, and visualize this trajectory in Fig.~\ref{fig.reasoning}. During the generation stage, each candidate formula is accompanied by a short rationale, which makes the LLM's hypothesis formation explicit. The initial proposal (depth 0) is a simple baseline in which the bulk modulus is treated as the dominant predictor and Poisson's ratio acts as a correction term. At depth 1, guided by evaluation feedback, the agent introduces an exponential-decay factor, reflecting the hypothesis that the effect of ductility on hardness is nonlinear rather than purely linear; this modification improves the score from 0.254 to 0.193. Across subsequent depths, the refinement operations can be broadly categorized as either \emph{mathematical} (e.g., adding free parameters or changing the decay functional form from depth 1 to 2 and 2 to 3) or \emph{physical} (introducing additional physically motivated variables). Notably, the final refinement (depth 3 to 4) incorporates the shear modulus into the expression, yielding a clear gain in predictive performance. This trajectory-level view provides a transparent account of how the ``Agent Theorist'' iteratively proposes, tests, and consolidates candidate material laws.
 
\begin{table*}[ht]
  \centering
  \caption{Average predictive performance of \textit{DeepSeek-V3} and \textit{Gemini-2.5-flash} across the considered tasks. Metrics are reported as ``train / test'' for the root-mean-square error (RMSE) and coefficient of determination ($R^2$). Best values are highlighted in \emph{italic}.}
  \label{tab.performance}
  \begin{tabular}{ccccc}
  \toprule
  \multirow{2}{*}{Task} 
  & \multicolumn{2}{c}{DeepSeek-V3} 
  & \multicolumn{2}{c}{Gemini-2.5-flash} \\
  & RMSE & $R^2$ & RMSE & $R^2$ \\
  \midrule
  GFA ($\log [\mathrm{K/s}]$)      & 2.43 / 2.41 & 0.80 / 0.82 & \emph{1.92 / 1.52} & \emph{0.89 / 0.94} \\
  Hardness (GPa)                   & 5.05 / 4.86 & 0.84 / 0.84 & \emph{4.90 / 4.52} & \emph{0.85 / 0.86} \\
  Young's modulus (GPa)            & 40.73 / 39.86 & 0.72 / 0.73\footnote{For the Young's modulus task, one of the three \textit{DeepSeek-V3} runs failed to produce a competitive formula, leading to a markedly lower average. The three runs yielded $R^2$ values of 0.94/0.95, 0.36/0.36, and 0.87/0.89 (train/test), respectively.} & \emph{21.50 / 20.75} & \emph{0.94 / 0.94} \\
  \bottomrule
  \end{tabular}
\end{table*}

\subsection*{Formula performance}

Table~\ref{tab.performance} reports the predictive accuracy of the two LLM backbones, quantified by the root-mean-square error (RMSE) and coefficient of determination ($R^2$), on three representative tasks. Across all datasets, \textit{Gemini-2.5-flash} consistently outperforms \textit{DeepSeek-V3}, yielding RMSE reductions of  $\sim$5-50\%. Unlike synthetic benchmarks with negligible noise, materials datasets typically originate from experiments or simulations and therefore contain intrinsic uncertainty. This is evident in the Vickers hardness task, where measurements are experimentally derived; while the resulting $R^2$ values do not reach near-perfect levels ($<0.95$), the average performance ($R^2$$ \approx$0.85) is comparable to prior studies on the same dataset~\cite{tantardiniMaterialHardnessDescriptor2024}, indicating that the discovered expressions capture meaningful structure-property relationships. For Young's modulus, where labels come from first-principles calculations and are more internally consistent, both models achieve substantially higher accuracy, with $R^2$ approaching 0.95 (with one failed run as an exception). Overall, these results suggest that the proposed framework is robust across both experiment- and simulation-based settings. Given the superior test-set performance of formulas generated by \textit{Gemini-2.5-flash}, we further present the corresponding regression plots in supplementary Fig.~S7 and interpret the discovered expressions from a physical perspective.

\emph{Glass-forming ability (GFA)}. Despite being supplied with more than twenty candidate descriptors, the highest-accuracy expression selected by our framework is remarkably compact and relies only on three characteristic temperatures: the glass transition temperature $T_g$, the crystallization temperature $T_x$, and the liquidus temperature $T_l$. The resulting formula for the logarithm of the critical cooling rate $R_\text{c}$ is

\begin{equation}\label{eq.gfa}
	\log (R_\text{c}) = -0.015\cdot\left(0.76\frac{T_\text{g}}{T_\text{l}}-\frac{T_\text{g}}{T_\text{x}}\right)-0.42.
\end{equation}

During formula generation, we explicitly discouraged the LLMs from combining all available descriptors and instead prompted them to select a small subset with clear physical relevance, guided by both prior knowledge and retrieved domain information. Notably, many widely used empirical GFA criteria in the literature are formulated primarily in terms of $\Tg$, $\Tx$, and $\Tl$. Consistent with this established practice, the LLMs preferentially selected these temperatures and recovered a compact, high-accuracy relationship that aligns with prevailing physical intuition.

The generated explanation from LLM to this formula is below:

\begin{quote}
\itshape
This formula emphasizes the balance between $T_\text{g}/T_\text{l}$ (enhancing GFA by stabilizing the undercooled liquid at higher temperatures) and $T_\text{g}/T_\text{x}$ (penalizing systems with low $\Tx$, which reduces the $\Delta T_\text{x}$ stability window). The term 0.76($T_\text{g}/T_\text{l}$) prioritizes a higher $T_\text{g}$ relative to $T_\text{l}$, delaying crystallization, while -($T_\text{g}/T_\text{x}$) discourages a small $\Delta T_\text{x}$, aligning with the context's emphasis on $\Delta T_\text{x}$ and temperature interplay. This combination effectively captures the trade-off between thermal stability and crystallization resistance, correlating with $\log R_\text{c}$ as improved GFA lowers the required cooling rate.
\end{quote}
These results suggest that the LLM does not explore the space of symbolic expressions arbitrarily; rather, it exploits embedded scientific knowledge to restrict the search to physically meaningful and interpretable forms. Guided by background information retrieved via RAG, Eq.~\eqref{eq.gfa} can be interpreted as a combination of two established dimensionless indicators: the reduced glass transition temperature $T_\text{g}/T_\text{l}$, originally proposed by Turnbull~\cite{turnbull_under_1969}, and the $T_\text{g}/T_\text{x}$ (or equivalently $\Delta T_\text{x}$) criterion~\cite{INOUE1993473}. The numerical prefactors ($-0.015$, $0.76$, and $-0.42$) are obtained by fitting. Overall, the expression couples thermodynamic driving forces for glass formation with kinetic resistance to crystallization, yielding a strong correlation with the critical cooling rate.

\emph{Vickers hardness}. For the hardness prediction task, the framework identifies three key elastic descriptors, the Voigt--Reuss--Hill (VRH) averaged bulk modulus $B_\text{VRH}$, VRH averaged shear modulus $G_\text{VRH}$, and Poisson's ratio $\nu$, to model the macroscopic Vickers hardness:
\begin{equation}
  H_\text{v} = 0.27\,B_\text{VRH}\frac{0.16}{0.03 + \nu} - 0.13\,G_\text{VRH}.
\end{equation}
The resulting form underscores the dominant role of elastic stiffness in governing hardness, with both $B_\text{VRH}$ and $G_\text{VRH}$ contributing explicitly. This is consistent with established empirical correlations (e.g., Chen's model~\cite{chen2011modeling}) that relate hardness to the combined effects of bulk and shear resistance to deformation.

\emph{Young’s modulus of MPEAs.} For the Young's modulus prediction task, the framework yields a compact expression in terms of four physically motivated descriptors: the average valence electron concentration ($\overline{VEC}$), electronegativity difference ($\Delta\chi$), average atomic mass ($\bar{m}$), and atomic-radius mismatch factor ($r_\gamma$). The resulting model reads
\begin{equation}
\begin{aligned}
E ={} & -38.27\,\overline{VEC} - 396.78\,\Delta\chi + 0.67\,\bar{m} \\
      & - 1102.52\,r_\gamma + 264.99\,(\overline{VEC}\times r_\gamma).
\end{aligned}
\end{equation}

Beyond the linear contributions, the inclusion of the interaction term $\overline{VEC}\times r_\gamma$ explicitly accounts for the coupling between electronic effects and atomic-size mismatch, i.e., how lattice distortion modulates the influence of valence electron concentration on stiffness. The resulting model attains consistently high predictive performance on both the training and test sets, with $R^2$ approaching $0.97$. As shown in the next section, the same expression also transfers well to previously unseen MPEA compositions, indicating strong generalization and promising extrapolative capability.

\emph{Human in the loop.} After deriving the closed-form expressions, we further refine them using basic physical insight and algebraic simplification to improve interpretability. For the Vickers hardness relation, the Poisson's ratio typically lies in the range of $\mathcal{O}(10^{-1})$. Consequently, the additive constant $0.03$ in the denominator is comparatively small and can be neglected to obtain a simpler, physically transparent form,
\begin{equation}
  H'=\text{param[1]}\frac{B_\text{VRH}}{\nu}+\text{param[2]}G_\text{VRH}
\end{equation}
The coefficients were re-optimized on the same training set, yielding $\text{param[1]}=0.02$ and $\text{param[2]}=0.03$. On the test set, this simplified model achieves $R^2=0.854$ with $\text{RMSE}=4.60\ \text{GPa}$. Although the RMSE increases by $\sim$5\% relative to the original expression (4.39~GPa), the resulting formula is substantially more compact and easier to interpret.

For the Young's modulus model of MPEAs, one may \emph{a priori} expect the average atomic mass $\bar{m}$ to play a limited role in elastic stiffness compared with electronic and size-mismatch descriptors. To test this hypothesis, we removed the $\bar{m}$ term and re-evaluated the resulting expression. This simplification led to a noticeable reduction in predictive performance on the test set, with $R^2$ decreasing from $0.97$ to $0.89$, indicating that $\bar{m}$ contributes non-negligible information within the present dataset and feature set. Overall, this ``human-in-the-loop'' step provides a practical mechanism to interrogate the physical plausibility of individual terms, quantify the accuracy-interpretability trade-off, and further refine the formulas produced by the multi-agent framework.

\section*{Application: Young's Modulus for MPEAs}

To assess the practical utility of the derived Young's modulus expression in alloy design, we apply it to compositionally more complex quaternary and quinary MPEA systems and benchmark its predictions against reliable atomistic-level simulation results. Notably, the formula was trained exclusively on data from binary and ternary alloys; therefore, this exercise provides a stringent test of its transferability to higher-order composition spaces.

Before deploying the formula for practical alloy design, we first validate its behavior on binary alloys contained in the training and test datasets. We consider four Nb-containing systems (Nb-Mo, Nb-Ta, Nb-V, and Nb-W) and compute the Young's modulus predicted by our formula as a function of Nb concentration, using DFT results as the reference. As shown in Fig.~S8, the predicted curves reproduce the overall DFT trends across composition for all systems. In the Nb-V alloy, the DFT data exhibit noticeable composition-dependent fluctuations that are not captured by the formula; this likely reflects the predominantly linear structure of the derived expression, which favors smooth, near-linear composition dependence. Quantitatively, the mean absolute percentage error (MAPE) across these systems is $12\%$-$15\%$, with $\mathrm{RMSE}=12$-$21$~GPa. Although Nb-V has a relatively low modulus and the model does not reproduce the fine-scale variations, its overall numerical error remains comparable to that of the other binaries. These results indicate that the formula provides a physically reasonable approximation for Nb-based binary alloys within the available dataset.


\begin{figure*}[ht]
	\centering
	\includegraphics[width=0.95\textwidth]{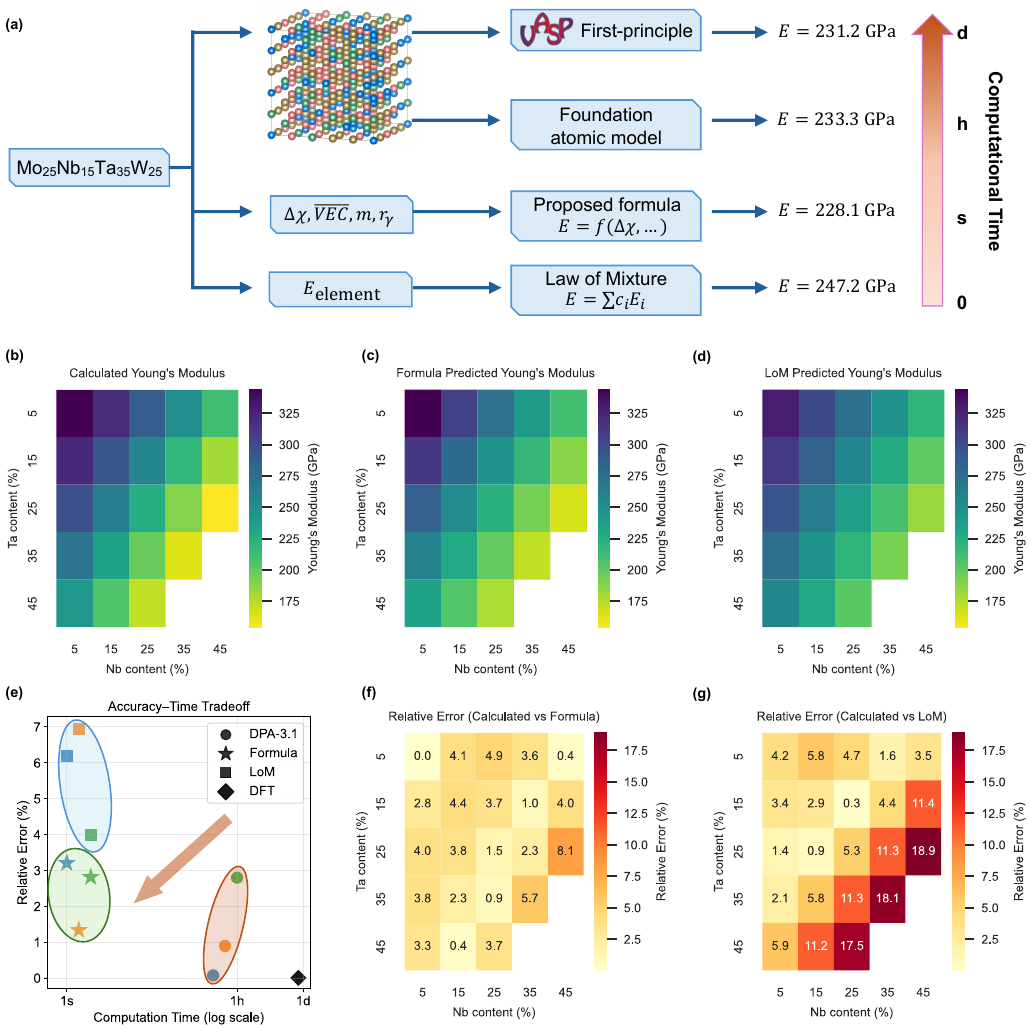}
	\caption{Comparison of Young's modulus predictions for the quaternary alloy $\text{Mo}_{25}\text{Ta}_{x}\text{Nb}_{y}\text{W}_{z}$ using DFT, the foundation atomic model DPA-3.1-3M, the proposed closed-form formula, and the law of mixtures (LoM). (a) Schematic workflow comparison from top to bottom: DFT, DPA-3.1-3M, proposed formula, and LoM. The last two columns report the predicted modulus for the representative alloy $\text{Mo}_{25}\text{Nb}_{15}\text{Ta}_{35}\text{W}_{25}$ and the corresponding computational cost, highlighting the favorable accuracy-efficiency trade-off of the proposed formula. (b) DPA-3.1-3M-predicted modulus map over the composition space. (c) Modulus map predicted by the proposed formula. (d) Modulus map predicted by LoM. (e) Accuracy versus computational cost for three representative compositions: Mo$_{25}$Nb$_5$W$_{25}$Ta$_{45}$, Mo$_{25}$Nb$_{15}$W$_{25}$Ta$_{35}$, and Mo$_{25}$Nb$_{25}$W$_{25}$Ta$_{25}$; DFT is taken as the reference (zero relative error). (f) Distribution of mean absolute percentage error (MAPE) of the proposed formula relative to DPA-3.1-3M across the sampled compositions. (g) MAPE distribution of LoM relative to DPA-3.1-3M.}
	\label{fig.dpa_mo}
\end{figure*}

\begin{table*}[]
	\centering
	\caption{Application of the proposed closed-form formulas for the quinary Mo-Nb-Ta-W-V alloy: comparison of predicted properties with atomistic results from the foundation model DPA-3.1-3M for representative optimized compositions.}
	\label{tab.quinary}
	\begin{tabular}{ccccccccc}
		\toprule
		& \multicolumn{2}{c}{DPA-3.1-3M} & &\multicolumn{2}{c}{Formula} & &\multicolumn{2}{c}{Error (\%)} \\
		Composition & E (GPa)          & B/G       && E (GPa)             & B/G         && E (GPa)           & B/G         \\
		\midrule
		$\text{Mo}_{24}\text{Nb}_{5}\text{Ta}_{33}\text{W}_{6}\text{V}_{32}$       & 158.63      & 3.57     && 166.73        & 3.51        && 5.10        & 1.55        \\
		$\text{Mo}_{24}\text{Nb}_{5}\text{Ta}_{32}\text{W}_{32}\text{V}_{7}$       & 253.71      & 2.53      && 245.90        & 2.72       && 3.08        & 7.54        \\
		$\text{Mo}_{35}\text{Nb}_{5}\text{Ta}_{19}\text{W}_{33}\text{V}_{8}$     & 281.22      & 2.33      && 270.85        & 2.53       && 3.69        & 8.63\\
		\bottomrule       
	\end{tabular}
\end{table*}

To probe the transferability of the formula to higher-dimensional composition spaces relevant to practical MPEA design, we further apply it to previously unseen Nb-Mo-Ta-W quaternary alloys. Specifically, for each of the four constituent elements, we construct a composition series by fixing its atomic fraction at 25~at.\% and varying the other three components to examine how the predicted Young's modulus evolves with composition. For example, in Fig.~\ref{fig.dpa_mo}, Mo is held constant at 25~at.\%, while the Nb and Ta contents are independently varied from 5 to 45~at.\% (with W adjusted to satisfy the compositional constraint), yielding a total of 22 distinct compositions in this series.

To enable rapid, high-throughput evaluation of these quaternary compositions, we compute elastic constants via molecular statics using a foundation atomic model with DFT-level accuracy. Specifically, we employ the DPA-3.1-3M potential~\cite{zhang2025graphneuralnetworkera}, whose \textit{alloy\_tongqi} branch is fine-tuned for alloy mechanical and defect properties. All calculations are executed using the cloud-native workflow Alloy Property EXplorer (APEX)~\cite{liAPEXAutomatedCloudnative2025a} in combination with LAMMPS. As shown in Supplementary Fig.~S9, DPA-3.1-3M reproduces DFT Young's modulus with deviations $<$3\%, providing a reliable atomistic benchmark for the present study. For each Nb-Mo-Ta-W composition, we construct a BCC random solid-solution supercell in LAMMPS and sample five independent random atomic configurations; the reported Young's modulus is the configuration-averaged value. To ensure consistency with the dataset used to derive the formula, the modulus is obtained from the calculated elastic constants using the Voigt-Reuss-Hill averaging scheme.

To demonstrate the predictive capability of the proposed formula, we benchmark its Young's modulus estimates against (i) atomistic results obtained with DPA-3.1-3M and (ii) the conventional law of mixtures (LoM) baseline. The LoM assumes a linear interpolation of the elemental modulus with composition,
\begin{equation}
  E = \sum_i c_i E_i
\end{equation}
where $c_i$ is the atomic fraction of element $i$ and $E_i$ is the Young's modulus of the corresponding pure element obtained from first-principles calculations.

Fig.~\ref{fig.dpa_mo} compares the Young's modulus of the quaternary series $\text{Mo}_{25}\text{Ta}_{x}\text{Nb}_{y}\text{W}_{z}$ ($5\leq x,y \leq 45$, with $z$ determined by the composition constraint) predicted by DPA-3.1-3M, the proposed formula, and the LoM baseline. Fig.~\ref{fig.dpa_mo}(a) further illustrates, for a representative composition, both the predicted modulus and the computation time required by first-principles calculations, the foundation atomic model, the proposed formula, and LoM. Atomistic approaches (DFT and DPA-3.1-3M) explicitly construct solid-solution supercells and therefore typically require $>$1 h per composition, whereas the proposed formula and LoM involve only numerical evaluation and complete within seconds. The DPA-3.1-3M results in Fig.~\ref{fig.dpa_mo}(b) reveal a clear compositional dependence: $E$ decreases with increasing Nb and Ta contents and increases with increasing W content. Importantly, the proposed formula reproduces the same trend across the entire composition space (Fig.~\ref{fig.dpa_mo}(c)), indicating that it captures the underlying physics governing the modulus variation. In contrast, LoM yields a qualitatively similar tendency (Fig.~\ref{fig.dpa_mo}(d)) but produces an overly smooth response, which obscures composition-dependent differences in $E$. Figs.~\ref{fig.dpa_mo}(f,g) quantify the accuracy using the mean absolute percentage error (MAPE) relative to DPA-3.1-3M. For the proposed formula, the MAPE is predominantly $<$5\%, with only two compositions in the 5-10\% range, demonstrating robust predictive accuracy throughout the quaternary space. LoM exhibits substantially larger errors, exceeding 17\% for certain compositions; overall, the proposed formula reduces the MAPE by up to 78.8\% (Mo$_{25}$Nb$_{25}$W$_{5}$Ta$_{45}$) relative to LoM. Similar accuracy ($<$5\% MAPE) is obtained for the other three 25~at.\% fixed-element series, with consistently reduced errors compared with LoM (Supplementary Figs.~S10-S12). Finally, Fig.~\ref{fig.dpa_mo}(e) summarizes the accuracy-efficiency trade-off using DFT as the reference for three representative alloys: Mo$_{25}$Nb$_5$W$_{25}$Ta$_{45}$, Mo$_{25}$Nb$_{15}$W$_{25}$Ta$_{35}$, and Mo$_{25}$Nb$_{25}$W$_{25}$Ta$_{25}$. Overall, the proposed multi-agent-derived formula offers the most favorable balance between computational cost and predictive accuracy among the methods considered.


We further extend the framework to the quinary Mo-Nb-Ta-W-V system, where the combinatorial composition space becomes prohibitively large for exhaustive enumeration. We therefore perform composition optimization using a genetic algorithm. Based on the same dataset and multi-agent workflow, we use the Gemini-2.5-flash model to derive closed-form expressions for the bulk modulus $B$ and shear modulus $G$:
\begin{equation}
  B = -97.58\overline{E_\text{coh}}+7.56\overline{E_\text{coh}}^2+98.55\overline{\text{VEC}} + 1.18S_\text{conf}
\end{equation}
\begin{equation}
  G = -107.39\overline{\text{VEC}} + 62.43\log(\overline{E_\text{coh}}) + 18.05\overline{\text{VEC}}^2
\end{equation}
where $\overline{E_\text{coh}}$ denotes the composition-averaged cohesive energy of the constituent elements, $\overline{\text{VEC}}$ is the averaged valence electron concentration, and $S_\text{conf}$ is the ideal configurational entropy. These models achieve $R^2=0.973$ for $B$ and $R^2=0.946$ for $G$ on the held-out test set. Because practical alloy development commonly involves competing objectives, we formulate a multi-objective design problem for the quinary alloys by simultaneously maximizing the Young's modulus $E$ and Pugh's ratio $B/G$. We solve this problem using the Non-dominated Sorting Genetic Algorithm II (NSGA-II), imposing the composition bounds $5\% \leq x \leq 35\%$ for each element $x\in\{\text{Mo, Nb, Ta, W, V}\}$ with a 1~at.\% step. We then select three representative compositions from the resulting Pareto front and validate the formula-based predictions against atomistic calculations. As summarized in Table~\ref{tab.quinary}, the predicted $E$ and $B/G$ agree well with DPA-3.1-3M results, with MAPE $<$10\% for both objectives. This quinary case study highlights the extrapolative capability of the derived formulas and their practical utility for rapid, multi-objective alloy design.

\section*{Discussion}
In this work, we investigated the capability of large language models (LLMs) to derive compact, interpretable scientific relationships from experimental and simulation datasets. With only minor changes to the task specification and input data, the proposed multi-agent framework was readily transferred across three representative materials problems: (i) glass-forming ability (GFA) of metallic glasses, (ii) Vickers hardness of inorganic compounds, and (iii) Young's modulus of multi-principal element alloys (MPEAs). In all cases, the resulting formulas reproduce the underlying measurements or reference calculations with high fidelity while remaining physically interpretable. Importantly, the LLM-based agents can provide expert-like reasoning about the inferred relations; for example, the RAG-assisted ``LLM scientist'' can contextualize and rationalize the GFA expression using established metallurgical principles rather than treating the model purely as a numerical fit.

Beyond interpolation within the available data, the derived expressions also exhibit strong extrapolation performance in the MPEA case. To explicitly test out-of-distribution generalization, we evaluated quaternary MPEAs that were not included in the training set. Relative to classical baselines such as the law of mixtures, the proposed formula yields substantially lower errors when benchmarked against results from the foundation atomic model (DPA-3.1-3M) and DFT. Moreover, a multi-objective composition optimization for quinary alloys further illustrates the practical utility of the formulas for alloy design, where both mechanical performance metrics and trade-offs must be considered simultaneously.

Methodologically, our approach combines the prior knowledge encoded in LLMs with structured search (e.g., beam search) and iterative feedback, enabling the discovery of explicit analytical expressions. In contrast to black-box predictors, which often provide limited mechanistic insight, the LLM-assisted pipeline produces transparent formulas that can be inspected, interpreted, and directly embedded into downstream workflows. This transparency makes the approach attractive for materials design, experimental data analysis, and the identification of physically meaningful descriptors. The consistent performance across distinct tasks supports the feasibility of LLM-driven scientific modeling and suggests a pathway toward increased automation in hypothesis generation, model derivation, and data-centric discovery.

Several limitations remain. First, despite broad pretraining, LLMs can lack reliable domain depth in specialized areas of physical science, particularly where the public training corpus is sparse. We partially address this issue via retrieval-augmented generation (RAG); however, because the framework currently lacks direct integration with specialized literature databases, data curation and literature selection are still semi-automated and require human input. This bottleneck motivates domain-specific retrieval infrastructures that tightly couple LLM agents with academic search engines and structured repositories for more autonomous knowledge acquisition. Second, symbolic regression is fundamentally a search problem; more efficient search strategies could reduce computational cost and improve solution quality. A promising direction is hybridization with conventional symbolic regression methods: LLMs propose physically plausible formula ``skeletons'', while dedicated optimization routines refine constants and functional forms to balance interpretability and accuracy. Third, our current selection procedure largely relies on goodness-of-fit metrics. As observed in the human-in-the-loop refinement, this can yield overly complex constants and formulas whose performance is sensitive to parameter values, thereby weakening physical interpretability. Developing richer selection criteria, for example, incorporating complexity penalties, dimensional/units consistency, robustness, and prior physical constraints, is an important next step.

Looking forward, multiple avenues can further improve the efficiency and scope of the framework. At present, the pipeline primarily operates on numerical tabular data, whereas scientific evidence is often multimodal, combining text, tables, and images or spectra (e.g., SEM micrographs, XRD patterns). Extending the framework to multimodal inputs would substantially broaden its applicability and enable direct use of primary characterization outputs. Ultimately, we envision a more fully automated scientific research platform in which the proposed multi-agent system supports hypothesis generation, theoretical derivation, and results interpretation. When integrated with simulation and experimental agents, such systems could evolve from passive analytical tools into active participants in accelerating materials discovery and advancing fundamental understanding.

\section*{Methods}

\subsection*{Beam search algorithm}
To efficiently explore the combinatorial space of candidate symbolic expressions, we employ a beam search strategy to construct and refine formulas. Exhaustive breadth-first or depth-first enumeration quickly becomes intractable because the number of possible expressions grows exponentially with formula depth. Beam search mitigates this issue by maintaining a fixed-size set (beam) of the $K$ highest-scoring candidates at each depth and expanding only these candidates in the next step. In this way, the search remains computationally bounded while preferentially allocating evaluations to promising regions of the expression space. The corresponding pseudo-code is provided in Algorithm~\ref{beam}.

\begin{algorithm}[H]
  \caption{Beam search}
  \label{beam}
  \begin{algorithmic}
    \State \textbf{Function:} Beam search
    \State \textbf{Input:} data $Db$, maximum depth $D$
    \State $beam \gets [\textit{root\_node}]$
    \For{$d = 1$ to $D$}
      \State $next\_beam \gets [\ ]$
      \ForAll{$node \in beam$}
        \If{$d == 1$}
          \State $Re \gets \textit{Background information}$
        \Else
          \State $Re \gets \textsc{Reflection\_Agent}(node.trajectory)$
        \EndIf
        \State $f \gets \textsc{Generation\_Agent}(node, Re)$
        \State $fc \gets \textsc{Evaluation\_Agent}(f, Db)$
        \State \textit{append} $fc$ \textit{to} $next\_beam$
      \EndFor
      \State \textit{keep top-}$K$ \textit{formulas in} $next\_beam$
      \State \textit{update} $best\_formula$, $best\_score$
      \State $beam \gets next\_beam$
    \EndFor
  \end{algorithmic}
\end{algorithm}

\subsection*{Agent setup}
All agents in our framework are implemented using the OpenAI SDK; the same design is directly portable to any OpenAI-compatible API. The overall workflow consists of (i) data preparation, (ii) multi-agent symbolic regression, and (iii) post hoc explanation of the selected formulas.

During data preparation, a \emph{data recommendation agent} proposes suitable independent variables and candidate operators for symbolic regression, assisted by retrieval-augmented generation (RAG) to ensure that the suggestions are grounded in domain knowledge. In addition, a \emph{data retrieval agent} queries elemental attributes (e.g., atomic and thermodynamic descriptors) for alloy systems via \textit{matminer}~\cite{ward2018matminer}. To ensure a fair comparison across different LLM backbones, the final set of independent variables and operator libraries is held fixed throughout all experiments.

The symbolic regression stage comprises three agents with complementary roles: (i) a \emph{generation agent} that proposes candidate symbolic expressions, (ii) an \emph{evaluation agent} that scores candidates based on their predictive performance and constraint compliance, and (iii) a \emph{reflection agent} that analyzes failure modes and provides targeted feedback to guide subsequent generations. All agents are instantiated using task-specific prompt templates; the full prompt details are provided in the Supplementary Information (SI). The prompt templates used by the \emph{explanation agent} in the interpretation stage are also included in the SI for completeness and reproducibility.

We implement RAG using LightRAG~\cite{guoLightRAGSimpleFast2025}, a lightweight GraphRAG system. In the experiments reported here, because the input datasets and candidate variables are fixed a priori, we populate the retrieval corpus with one representative reference per task that provides the definitions and physical motivations for the selected descriptors: \cite{guoIdentifyBestGlass2010} for GFA, \cite{tantardiniMaterialHardnessDescriptor2024} for hardness, and \cite{vazquezEfficientMachinelearningModel2022a} for Young's modulus. These references supply the background context required for grounded variable selection and scientifically coherent explanations.

\subsection*{Base models}
We evaluate the framework using two widely adopted LLM backbones: Gemini-2.5-flash and DeepSeek-V3. To enable a controlled comparison of symbolic regression performance, the candidate independent variables and operator set are held constant across all models and across repeated runs within each task. For each task-model pair, we execute the identical agent configuration (see SI for full settings) three independent times and report performance metrics averaged over these runs.

\subsection*{GFA of metallic glasses}
GFA quantifies the propensity of an alloy melt to bypass crystallization and form an amorphous solid under a given processing condition~\cite{liDatadrivenDiscoveryUniversal2022}. A widely used experimental proxy for GFA is the critical cooling rate $R_\text{c}$: alloys with lower $R_\text{c}$ exhibit higher GFA. Direct measurements of $R_\text{c}$, however, are typically time-consuming and costly, which motivates the development of predictive and interpretable descriptors. Prior studies have shown that quantitative GFA criteria can be constructed from characteristic temperatures obtained by thermal analysis, including the glass transition temperature $T_\text{g}$, the onset crystallization temperature $T_\text{x}$, and the liquidus temperature $T_\text{l}$. In this work, we use these experimentally accessible quantities as core variables, and further augment the feature set with composition-derived atomic descriptors, such as the valence electron concentration (VEC) and mixing enthalpy, computed using the \textit{matminer} package~\cite{ward2018matminer}. The objective of this task is to derive an explicit symbolic expression that predicts $\log(R_\text{c})$ as an interpretable measure of GFA. The dataset comprises 56 metallic-glass compositions compiled from the literature and is used for formula discovery.

\subsection*{Vickers hardness of materials}
Hardness is a key mechanical property that characterizes a material's resistance to permanent (plastic) deformation. Several standardized indentation-based protocols have been developed to quantify hardness, including the Rockwell, Brinell, and Vickers tests. Among these, Vickers hardness ($H_\mathrm{V}$) is widely adopted due to its high measurement precision and broad applicability across material classes, ranging from soft metals to ultrahard materials such as diamond~\cite{smith1922accurate,tu_2025_nc}. Despite its utility, experimental determination of $H_\mathrm{V}$ remains relatively costly and time-consuming, which limits rapid materials screening. To enable high-throughput discovery, this task aims to correlate macroscopic Vickers hardness with fundamental descriptors that can be obtained from first-principles calculations, including elastic moduli, Poisson's ratio, and atomic size-related features. We use the dataset curated by~\cite{TANTARDINI2024102402}, comprising 61 compounds with experimentally measured $H_\mathrm{V}$ and DFT-derived microscopic properties, as the basis for symbolic formula discovery.

\subsection*{Young's modulus of MPEAs}
Refractory multi-principal element alloys (MPEAs) are an emerging class of structural materials composed primarily of refractory elements (e.g., Ti, Nb, Ta, W, Mo, and Cr). Owing to their excellent mechanical performance at elevated temperatures, they are promising for applications in extreme environments, including aerospace, power generation, and nuclear systems. Young's modulus ($E$) of refractory MPEAs is typically high ($\sim$200--400~GPa), and application-specific design often requires tailoring $E$ in concert with other properties. Consequently, establishing composition-property relationships that enable efficient modulus tuning is of practical importance. Previous studies suggest that $E$ in MPEAs is strongly influenced by composition-derived atomic descriptors, such as the valence electron concentration and atomic size mismatch~\cite{khakurelMachineLearningAssisted2021}. In this task, we aim to derive an interpretable symbolic expression that links such atomic descriptors to DFT-predicted Young's modulus. We use the DFT dataset reported by~\cite{vazquezEfficientMachinelearningModel2022a}, which focuses on Nb-W-Mo-Ta-V refractory MPEAs and provides elastic constants for 170 binary and ternary compositions. Atomic-level descriptors are computed from composition using the \textit{matminer}~\cite{ward2018matminer} Python package.

\subsection*{Genetic algorithm for quinary alloy composition optimization}
The compositional design space of quinary alloys is high dimensional and prohibitively large for exhaustive enumeration. We therefore performed multi-objective optimization using the NSGA-II genetic algorithm as implemented in the \textit{pymoo}~\cite{pymoo} framework to simultaneously maximize the $B/G$ ratio (taken as a proxy for ductility) and Young's modulus $E$ (stiffness). The decision variables were the atomic fractions of Mo, Nb, Ta, and W, with the V fraction determined by mass balance, i.e., $x_\mathrm{V}=1-(x_\mathrm{Mo}+x_\mathrm{Nb}+x_\mathrm{Ta}+x_\mathrm{W})$. Each elemental fraction was constrained to 5-35~at.\%, ensuring physically realistic compositions and avoiding extreme corner solutions. For each candidate composition, the bulk modulus $B$, shear modulus $G$, and Young's modulus $E$ were evaluated using the physically interpretable formulas developed in this work, and the resulting $B/G$ and $E$ values served directly as the two objective functions. The NSGA-II procedure used integer encoding, a population size of 100, and was evolved for 50 generations. Candidate solutions were ranked via non-dominated sorting, and diversity along the front was maintained using the crowding-distance criterion. The final non-dominated set constitutes the Pareto-optimal front, which explicitly captures the stiffness-ductility trade-off across the quinary alloy design space.

\section*{Data availability}
The dataset used in the metallic-glass case study (56 samples with characteristic temperatures and critical cooling rates) is provided in the Supplementary Information. All other datasets analyzed in this work were obtained from the original publications cited in the corresponding sections and can be accessed through those references.

\section*{Code availability}
The source code developed for data processing, model construction, and optimization in this study will be deposited in a public GitHub repository upon acceptance of the manuscript. The code is also available from the authors upon reasonable request.

\def\bibsection{\section*{\refname}}
\bibliography{sn-bibliography.bib}

@inproceedings{ICSR,
    title = "In-Context Symbolic Regression: Leveraging Large Language Models for Function Discovery",
    author = "Merler, Matteo  and
      Haitsiukevich, Katsiaryna  and
      Dainese, Nicola  and
      Marttinen, Pekka",
    editor = "Fu, Xiyan  and
      Fleisig, Eve",
    booktitle = "Proceedings of the 62nd Annual Meeting of the Association for Computational Linguistics (Volume 4: Student Research Workshop)",
    month = aug,
    year = "2024",
    address = "Bangkok, Thailand",
    publisher = "Association for Computational Linguistics",
    url = "https://aclanthology.org/2024.acl-srw.49",
    doi = "10.18653/v1/2024.acl-srw.49",
    pages = "427--444",
}

@ARTICLE{pymoo,
    author={J. {Blank} and K. {Deb}},
    journal={IEEE Access},
    title={pymoo: Multi-Objective Optimization in Python},
    year={2020},
    volume={8},
    number={},
    pages={89497-89509},
}

@inproceedings{Reflextion,
 author = {Shinn, Noah and Cassano, Federico and Gopinath, Ashwin and Narasimhan, Karthik and Yao, Shunyu},
 booktitle = {Advances in Neural Information Processing Systems},
 editor = {A. Oh and T. Naumann and A. Globerson and K. Saenko and M. Hardt and S. Levine},
 pages = {8634--8652},
 publisher = {Curran Associates, Inc.},
 title = {Reflexion: language agents with verbal reinforcement learning},
 volume = {36},
 year = {2023},
url = {https://proceedings.neurips.cc/paper_files/paper/2023/hash/1b44b878bb782e6954cd888628510e90-Abstract-Conference.html},
}

@article{turnbull_under_1969,
	title = {Under what conditions can a glass be formed?},
	volume = {10},
	issn = {0010-7514, 1366-5812},
	url = {http://www.tandfonline.com/doi/abs/10.1080/00107516908204405},
	doi = {10.1080/00107516908204405},
	number = {5},
	journal = {Contemporary Physics},
	author = {Turnbull, David},
	month = sep,
	year = {1969},
	pages = {473--488},
}

@article{INOUE1993473,
  title = {Glass-Forming Ability of Alloys},
  author = {Inoue, Akihisa and Zhang, Tao and Masumoto, Tsuyoshi},
  year = {1993},
  journal = {Journal of Non-Crystalline Solids},
  volume = {156--158},
  pages = {473--480},
  issn = {0022-3093},
  doi = {10.1016/0022-3093(93)90003-G}
}

@article{ward2018matminer,
  title={Matminer: An open source toolkit for materials data mining},
  author={Ward, Logan and Dunn, Alexander and Faghaninia, Alireza and Zimmermann, Nils ER and Bajaj, Saurabh and Wang, Qi and Montoya, Joseph and Chen, Jiming and Bystrom, Kyle and Dylla, Maxwell and others},
  journal={Computational Materials Science},
  volume={152},
  pages={60--69},
  year={2018},
  publisher={Elsevier}
}

@inproceedings{NEURIPS2020_6b493230,
 author = {Lewis, Patrick and Perez, Ethan and Piktus, Aleksandra and Petroni, Fabio and Karpukhin, Vladimir and Goyal, Naman and K\"{u}ttler, Heinrich and Lewis, Mike and Yih, Wen-tau and Rockt\"{a}schel, Tim and Riedel, Sebastian and Kiela, Douwe},
 booktitle = {Advances in Neural Information Processing Systems},
 editor = {H. Larochelle and M. Ranzato and R. Hadsell and M.F. Balcan and H. Lin},
 pages = {9459--9474},
 publisher = {Curran Associates, Inc.},
 title = {Retrieval-Augmented Generation for Knowledge-Intensive NLP Tasks},
 url = {https://proceedings.neurips.cc/paper_files/paper/2020/file/6b493230205f780e1bc26945df7481e5-Paper.pdf},
 volume = {33},
 year = {2020}
}

@misc{guoLightRAGSimpleFast2025,
  title = {{{LightRAG}}: {{Simple}} and {{Fast Retrieval-Augmented Generation}}},
  shorttitle = {{{LightRAG}}},
  author = {Guo, Zirui and Xia, Lianghao and Yu, Yanhua and Ao, Tu and Huang, Chao},
  year = 2025,
  month = apr,
  number = {arXiv:2410.05779},
  eprint = {2410.05779},
  primaryclass = {cs},
  publisher = {arXiv},
  doi = {10.48550/arXiv.2410.05779},
  urldate = {2025-10-23},
  archiveprefix = {arXiv}
}

@article{khakurelMachineLearningAssisted2021,
  title = {Machine Learning Assisted Prediction of the {{Young}}'s Modulus of Compositionally Complex Alloys},
  author = {Khakurel, Hrishabh and Taufique, M. F. N. and Roy, Ankit and Balasubramanian, Ganesh and Ouyang, Gaoyuan and Cui, Jun and Johnson, Duane D. and Devanathan, Ram},
  year = 2021,
  month = aug,
  journal = {Scientific Reports},
  volume = {11},
  number = {1},
  pages = {17149},
  issn = {2045-2322},
  doi = {10.1038/s41598-021-96507-0},
  urldate = {2025-10-23},
  langid = {english}
}

@article{chen2011modeling,
  title={Modeling hardness of polycrystalline materials and bulk metallic glasses},
  author={Chen, Xing-Qiu and Niu, Haiyang and Li, Dianzhong and Li, Yiyi},
  journal={Intermetallics},
  volume={19},
  number={9},
  pages={1275--1281},
  year={2011},
  publisher={Elsevier}
}

@article{liAPEXAutomatedCloudnative2025a,
  title = {{{APEX}}: An Automated Cloud-Native Material Property Explorer},
  author = {Li, Zhuoyuan and Wen, Tongqi and Zhang, Yuzhi and Liu, Xinzijian and Zhang, Chengqian and Pattamatta, A. S. L. Subrahmanyam and Gong, Xiaoguo and Ye, Beilin and Wang, Han and Zhang, Linfeng and Srolovitz, David J.},
  year = 2025,
  month = apr,
  journal = {npj Computational Materials},
  volume = {11},
  number = {1},
  pages = {88},
  issn = {2057-3960},
  doi = {10.1038/s41524-025-01580-y}
}

@article{guoIdentifyBestGlass2010,
  title = {Identify the Best Glass Forming Ability Criterion},
  author = {Guo, Sheng and Lu, Z. P. and Liu, C. T.},
  year = 2010,
  month = may,
  journal = {Intermetallics},
  volume = {18},
  number = {5},
  pages = {883--888},
  issn = {0966-9795},
  doi = {10.1016/j.intermet.2009.12.025},
  urldate = {2025-02-13}
}

@misc{zhang2025graphneuralnetworkera,
      title={A Graph Neural Network for the Era of Large Atomistic Models}, 
      author={Duo Zhang and Anyang Peng and Chun Cai and Wentao Li and Yuanchang Zhou and Jinzhe Zeng and Mingyu Guo and Chengqian Zhang and Bowen Li and Hong Jiang and Tong Zhu and Weile Jia and Linfeng Zhang and Han Wang},
      year={2025},
      eprint={2506.01686},
      archivePrefix={arXiv},
      primaryClass={physics.comp-ph},
      url={https://arxiv.org/abs/2506.01686}, 
}

@article{smith1922accurate,
  title={An accurate method of determining the hardness of metals, with particular reference to those of a high degree of hardness},
  author={Smith, Robert L and Sandly, GE},
  journal={Proceedings of the Institution of Mechanical Engineers},
  volume={102},
  number={1},
  pages={623--641},
  year={1922},
  publisher={SAGE Publications Sage UK: London, England}
}

@article{udrescuAIFeynmanPhysicsinspired2020,
  title = {{{AI Feynman}}: {{A}} Physics-Inspired Method for Symbolic Regression},
  shorttitle = {{{AI Feynman}}},
  author = {Udrescu, Silviu-Marian and Tegmark, Max},
  year = {2020},
  month = apr,
  journal = {Science Advances},
  volume = {6},
  number = {16},
  pages = {eaay2631},
  publisher = {American Association for the Advancement of Science},
  doi = {10.1126/sciadv.aay2631},
  langid = {american}
}

@article{jainCommentaryMaterialsProject2013,
  title = {Commentary: {{The Materials Project}}: {{A}} Materials Genome Approach to Accelerating Materials Innovation},
  shorttitle = {Commentary},
  author = {Jain, Anubhav and Ong, Shyue Ping and Hautier, Geoffroy and Chen, Wei and Richards, William Davidson and Dacek, Stephen and Cholia, Shreyas and Gunter, Dan and Skinner, David and Ceder, Gerbrand and Persson, Kristin A.},
  year = {2013},
  month = jul,
  journal = {APL Materials},
  volume = {1},
  number = {1},
  pages = {011002},
  issn = {2166-532X},
  doi = {10.1063/1.4812323},
  langid = {english}
}

@article{curtaroloAFLOWAutomaticFramework2012,
  title = {{{AFLOW}}: {{An}} Automatic Framework for High-Throughput Materials Discovery},
  shorttitle = {{{AFLOW}}},
  author = {Curtarolo, Stefano and Setyawan, Wahyu and Hart, Gus L. W. and Jahnatek, Michal and Chepulskii, Roman V. and Taylor, Richard H. and Wang, Shidong and Xue, Junkai and Yang, Kesong and Levy, Ohad and Mehl, Michael J. and Stokes, Harold T. and Demchenko, Denis O. and Morgan, Dane},
  year = {2012},
  month = jun,
  journal = {Computational Materials Science},
  volume = {58},
  pages = {218--226},
  issn = {0927-0256},
  doi = {10.1016/j.commatsci.2012.02.005}
}

@article{gossettAFLOWMLRESTfulAPI2018,
  title = {{{AFLOW-ML}}: {{A RESTful API}} for Machine-Learning Predictions of Materials Properties},
  shorttitle = {{{AFLOW-ML}}},
  author = {Gossett, Eric and Toher, Cormac and Oses, Corey and Isayev, Olexandr and Legrain, Fleur and Rose, Frisco and Zurek, Eva and Carrete, Jes{\'u}s and Mingo, Natalio and Tropsha, Alexander and Curtarolo, Stefano},
  year = {2018},
  month = sep,
  journal = {Computational Materials Science},
  volume = {152},
  pages = {134--145},
  issn = {0927-0256},
  doi = {10.1016/j.commatsci.2018.03.075}
}

@article{yangMatCloudHighthroughputComputational2018,
  title = {{{MatCloud}}: {{A}} High-Throughput Computational Infrastructure for Integrated Management of Materials Simulation, Data and Resources},
  shorttitle = {{{MatCloud}}},
  author = {Yang, Xiaoyu and Wang, Zongguo and Zhao, Xushan and Song, Jianlong and Zhang, Mingming and Liu, Haidong},
  year = {2018},
  month = apr,
  journal = {Computational Materials Science},
  volume = {146},
  pages = {319--333},
  issn = {0927-0256},
  doi = {10.1016/j.commatsci.2018.01.039}
}

@article{zagoracRecentDevelopmentsInorganic2019,
  title = {Recent Developments in the {{Inorganic Crystal Structure Database}}: Theoretical Crystal Structure Data and Related Features},
  shorttitle = {Recent Developments in the {{Inorganic Crystal Structure Database}}},
  author = {Zagorac, D. and M{\"u}ller, H. and Ruehl, S. and Zagorac, J. and Rehme, S.},
  year = {2019},
  month = oct,
  journal = {Journal of Applied Crystallography},
  volume = {52},
  number = {5},
  pages = {918--925},
  issn = {1600-5767},
  doi = {10.1107/S160057671900997X},
  langid = {english}
}

@article{taoMachineLearningPerovskite2021,
  title = {Machine Learning for Perovskite Materials Design and Discovery},
  author = {Tao, Qiuling and Xu, Pengcheng and Li, Minjie and Lu, Wencong},
  year = {2021},
  month = jan,
  journal = {npj Computational Materials},
  volume = {7},
  number = {1},
  pages = {1--18},
  publisher = {Nature Publishing Group},
  issn = {2057-3960},
  doi = {10.1038/s41524-021-00495-8},
  copyright = {2021 The Author(s)},
  langid = {english}
}

@article{jhaElemNetDeepLearning2018a,
  title = {{{ElemNet}}: {{Deep Learning}} the {{Chemistry}} of {{Materials From Only Elemental Composition}}},
  shorttitle = {{{ElemNet}}},
  author = {Jha, Dipendra and Ward, Logan and Paul, Arindam and Liao, Wei-keng and Choudhary, Alok and Wolverton, Chris and Agrawal, Ankit},
  year = {2018},
  month = dec,
  journal = {Scientific Reports},
  volume = {8},
  number = {1},
  pages = {17593},
  publisher = {Nature Publishing Group},
  issn = {2045-2322},
  doi = {10.1038/s41598-018-35934-y},
  copyright = {2018 The Author(s)},
  langid = {english}
}

@article{angelisArtificialIntelligencePhysical2023,
  title = {Artificial {{Intelligence}} in {{Physical Sciences}}: {{Symbolic Regression Trends}} and {{Perspectives}}},
  shorttitle = {Artificial {{Intelligence}} in {{Physical Sciences}}},
  author = {Angelis, Dimitrios and Sofos, Filippos and Karakasidis, Theodoros E.},
  year = {2023},
  month = jul,
  journal = {Archives of Computational Methods in Engineering},
  volume = {30},
  number = {6},
  pages = {3845--3865},
  issn = {1886-1784},
  doi = {10.1007/s11831-023-09922-z},
  langid = {english}
}

@article{wangScientificDiscoveryAge2023,
  title = {Scientific Discovery in the Age of Artificial Intelligence},
  author = {Wang, Hanchen and Fu, Tianfan and Du, Yuanqi and Gao, Wenhao and Huang, Kexin and Liu, Ziming and Chandak, Payal and Liu, Shengchao and Van Katwyk, Peter and Deac, Andreea and Anandkumar, Anima and Bergen, Karianne and Gomes, Carla P. and Ho, Shirley and Kohli, Pushmeet and Lasenby, Joan and Leskovec, Jure and Liu, Tie-Yan and Manrai, Arjun and Marks, Debora and Ramsundar, Bharath and Song, Le and Sun, Jimeng and Tang, Jian and Veli{\v c}kovi{\'c}, Petar and Welling, Max and Zhang, Linfeng and Coley, Connor W. and Bengio, Yoshua and Zitnik, Marinka},
  year = {2023},
  month = aug,
  journal = {Nature},
  volume = {620},
  number = {7972},
  pages = {47--60},
  publisher = {Nature Publishing Group},
  issn = {1476-4687},
  doi = {10.1038/s41586-023-06221-2},
  copyright = {2023 Springer Nature Limited},
  langid = {english}
}

@misc{duLLM4EDLargeLanguage2024,
  title = {{{LLM4ED}}: {{Large Language Models}} for {{Automatic Equation Discovery}}},
  shorttitle = {{{LLM4ED}}},
  author = {Du, Mengge and Chen, Yuntian and Wang, Zhongzheng and Nie, Longfeng and Zhang, Dongxiao},
  year = {2024},
  month = jul,
  number = {arXiv:2405.07761},
  eprint = {2405.07761},
  publisher = {arXiv},
  archiveprefix = {arXiv},
  langid = {english}
}

@misc{shojaeeLLMSRScientificEquation2024,
  title = {{{LLM-SR}}: {{Scientific Equation Discovery}} via {{Programming}} with {{Large Language Models}}},
  shorttitle = {{{LLM-SR}}},
  author = {Shojaee, Parshin and Meidani, Kazem and Gupta, Shashank and Farimani, Amir Barati and Reddy, Chandan K.},
  year = {2024},
  month = jun,
  number = {arXiv:2404.18400},
  eprint = {2404.18400},
  publisher = {arXiv},
  archiveprefix = {arXiv},
  langid = {english}
}

@inproceedings{NEURIPS2023_a3621ee9,
 author = {Li, Guohao and Hammoud, Hasan and Itani, Hani and Khizbullin, Dmitrii and Ghanem, Bernard},
 booktitle = {Advances in Neural Information Processing Systems},
 editor = {A. Oh and T. Naumann and A. Globerson and K. Saenko and M. Hardt and S. Levine},
 pages = {51991--52008},
 publisher = {Curran Associates, Inc.},
 title = {CAMEL: Communicative Agents for "Mind" Exploration of Large Language Model Society},
 url = {https://proceedings.neurips.cc/paper_files/paper/2023/file/a3621ee907def47c1b952ade25c67698-Paper-Conference.pdf},
 volume = {36},
 year = {2023}
}

@misc{gouCRITICLargeLanguage2024,
  title = {{{CRITIC}}: {{Large Language Models Can Self-Correct}} with {{Tool-Interactive Critiquing}}},
  shorttitle = {{{CRITIC}}},
  author = {Gou, Zhibin and Shao, Zhihong and Gong, Yeyun and Shen, Yelong and Yang, Yujiu and Duan, Nan and Chen, Weizhu},
  year = {2024},
  month = feb,
  number = {arXiv:2305.11738},
  eprint = {2305.11738},
  publisher = {arXiv},
  archiveprefix = {arXiv},
  langid = {english}
}

@article{liDatadrivenDiscoveryUniversal2022,
  title = {Data-Driven Discovery of a Universal Indicator for Metallic Glass Forming Ability},
  author = {Li, Ming-Xing and Sun, Yi-Tao and Wang, Chao and Hu, Li-Wei and Sohn, Sungwoo and Schroers, Jan and Wang, Wei-Hua and Liu, Yan-Hui},
  year = {2022},
  month = feb,
  journal = {Nature Materials},
  volume = {21},
  number = {2},
  pages = {165--172},
  publisher = {Nature Publishing Group},
  issn = {1476-4660},
  doi = {10.1038/s41563-021-01129-6},
  copyright = {2021 The Author(s), under exclusive licence to Springer Nature Limited},
  langid = {english}
}

@article{TANTARDINI2024102402,
  title = {Material Hardness Descriptor Derived by Symbolic Regression},
  author = {Tantardini, Christian and Zakaryan, Hayk A. and Han, Zhong-Kang and Altalhi, Tariq and Levchenko, Sergey V. and Kvashnin, Alexander G. and Yakobson, Boris I.},
  year = {2024},
  journal = {Journal of Computational Science},
  volume = {82},
  pages = {102402},
  issn = {1877-7503},
  doi = {10.1016/j.jocs.2024.102402}
}

@article{zeniGenerativeModelInorganic2025,
  title = {A Generative Model for Inorganic Materials Design},
  author = {Zeni, Claudio and Pinsler, Robert and Z{\"u}gner, Daniel and Fowler, Andrew and Horton, Matthew and Fu, Xiang and Wang, Zilong and Shysheya, Aliaksandra and Crabb{\'e}, Jonathan and Ueda, Shoko and Sordillo, Roberto and Sun, Lixin and Smith, Jake and Nguyen, Bichlien and Schulz, Hannes and Lewis, Sarah and Huang, Chin-Wei and Lu, Ziheng and Zhou, Yichi and Yang, Han and Hao, Hongxia and Li, Jielan and Yang, Chunlei and Li, Wenjie and Tomioka, Ryota and Xie, Tian},
  year = 2025,
  month = jan,
  journal = {Nature},
  pages = {1--3},
  publisher = {Nature Publishing Group},
  issn = {1476-4687},
  doi = {10.1038/s41586-025-08628-5},
  urldate = {2025-03-01},
  copyright = {2025 The Author(s), under exclusive licence to Springer Nature Limited},
  langid = {english}
}

@misc{niMatPilotLLMenabledAI2024,
  title = {{{MatPilot}}: An {{LLM-enabled AI Materials Scientist}} under the {{Framework}} of {{Human-Machine Collaboration}}},
  shorttitle = {{{MatPilot}}},
  author = {Ni, Ziqi and Li, Yahao and Hu, Kaijia and Han, Kunyuan and Xu, Ming and Chen, Xingyu and Liu, Fengqi and Ye, Yicong and Bai, Shuxin},
  year = 2024,
  month = nov,
  number = {arXiv:2411.08063},
  eprint = {2411.08063},
  primaryclass = {physics},
  publisher = {arXiv},
  doi = {10.48550/arXiv.2411.08063},
  urldate = {2025-10-23},
  archiveprefix = {arXiv}
}

@book{kuhn1997structure,
  title={The structure of scientific revolutions},
  author={Kuhn, Thomas S},
  volume={962},
  year={1997},
  publisher={University of Chicago press Chicago}
}

@inproceedings{
wu2024autogen,
title={AutoGen: Enabling Next-Gen {LLM} Applications via Multi-Agent Conversations},
author={Qingyun Wu and Gagan Bansal and Jieyu Zhang and Yiran Wu and Beibin Li and Erkang Zhu and Li Jiang and Xiaoyun Zhang and Shaokun Zhang and Jiale Liu and Ahmed Hassan Awadallah and Ryen W White and Doug Burger and Chi Wang},
booktitle={First Conference on Language Modeling},
year={2024},
url={https://openreview.net/forum?id=BAakY1hNKS}
}

@article{tantardiniMaterialHardnessDescriptor2024,
  title = {Material {{Hardness Descriptor Derived}} by {{Symbolic Regression}}},
  author = {Tantardini, Christian and Zakaryan, Hayk A. and Han, Zhong-Kang and Altalhi, Tariq and Levchenko, Sergey V. and Kvashnin, Alexander G. and Yakobson, Boris I.},
  year = 2024,
  month = oct,
  journal = {Journal of Computational Science},
  volume = {82},
  eprint = {2304.12880},
  primaryclass = {cond-mat},
  pages = {102402},
  issn = {18777503},
  doi = {10.1016/j.jocs.2024.102402},
  urldate = {2025-06-30},
  archiveprefix = {arXiv}
}

@article{vazquezEfficientMachinelearningModel2022a,
  title = {Efficient Machine-Learning Model for Fast Assessment of Elastic Properties of High-Entropy Alloys},
  author = {Vazquez, Guillermo and Singh, Prashant and Sauceda, Daniel and Couperthwaite, Richard and Britt, Nicholas and Youssef, Khaled and Johnson, Duane D. and Arr{\'o}yave, Raymundo},
  year = 2022,
  month = jun,
  journal = {Acta Materialia},
  volume = {232},
  pages = {117924},
  issn = {13596454},
  doi = {10.1016/j.actamat.2022.117924},
  urldate = {2025-08-11},
  langid = {english}
}

@article{liDeeplearningDensityFunctional2022,
  title = {Deep-Learning Density Functional Theory {{Hamiltonian}} for Efficient Ab Initio Electronic-Structure Calculation},
  author = {Li, He and Wang, Zun and Zou, Nianlong and Ye, Meng and Xu, Runzhang and Gong, Xiaoxun and Duan, Wenhui and Xu, Yong},
  year = 2022,
  month = jun,
  journal = {Nature Computational Science},
  volume = {2},
  number = {6},
  pages = {367--377},
  publisher = {Nature Publishing Group},
  issn = {2662-8457},
  doi = {10.1038/s43588-022-00265-6},
  urldate = {2025-03-01},
  copyright = {2022 The Author(s)},
  langid = {english}
}

@article{zhang_2025_nature,
	title={A multimodal robotic platform for multi-element electrocatalyst discovery},
	author={Zhang, Zhen and Ren, Zhichu and Hsu, Chia-Wei and Chen, Weibin and Hong, Zhang-Wei and Lee, Chi-Feng and Penn, Aubrey and Xu, Hongbin and Zheng, Daniel J and Miao, Shuhan and others},
	journal={Nature},
	pages={1--3},
	year={2025},
	publisher={Nature Publishing Group UK London},
	doi={10.1038/s41586-025-09640-5}
}

@article{tu_2025_nc,
	title = {Inch-Scale Ultrahard Diamond Wafer with 200 {{GPa}} Hardness via High-Frequency Pulsed Local Non-Equilibrium Growth},
	author = {Tu, Juping and Li, Jiayi and Wang, Yong and Zhao, Yun and Liu, Jinlong and Wei, Junjun and Chen, Liangxian and Zhang, Jianjun and Lu, Yang and Li, Chengming},
	year = 2025,
	month = dec,
	journal = {Nature Communications},
	publisher = {Nature Publishing Group},
	issn = {2041-1723},
	doi = {10.1038/s41467-025-66456-7}
}

\section*{Acknowledgments}
The work described is partially supported by a grant from the NSFC/RGC Joint Research
Scheme sponsored by the Research Grants Council of the Hong Kong Special Administrative Region, China and the National Natural Science Foundation of China (Project No. N\_HKU767/25). The authors would like to thank for startup funding from Materials Innovation Institute for Life Sciences and Energy (MILES), HKU-SIRI in Shenzhen for support of this manuscript. This work is partially supported by Research Grants Council, Hong Kong SAR through the General Research Fund (17210723, 17200424). T. W. acknowledges additional support by The University of Hong Kong (HKU) via seed funds (2509100468) and Guangdong Natural Science Fund (2025A1515012129). Y. L. acknowledges the supports from RGC under RFS2021-1S05 and from NSFC under 12525205.

\section*{Competing interests}
The authors declare no competing interests.

\section*{Supplementary Information}
Supplementary Notes

Supplementary Figures S1-12

Supplementary Tables S1-3

\end{document}